\documentclass{article}
\usepackage{makeidx}
\usepackage{graphicx}
\usepackage{amsmath}
\usepackage{amsfonts}
\usepackage{amssymb}

\begin{document}

\title{Backwards-induction outcome in a quantum game. }
\author{A. Iqbal and A.H. Toor\\Electronics Department, Quaid-i-Azam University, \\Islamabad, Pakistan\\email: qubit@isb.paknet.com.pk\\}
\maketitle
\begin{abstract}
In economics duopoly is a market dominated by two firms large enough to
influence the market price. Stackelberg presented a dynamic form of duopoly
that is also called `leader-follower' model. We give a quantum perspective on
Stackelberg duopoly that gives a backwards-induction outcome same as the Nash
equilibrium in static form of duopoly also known as Cournot's duopoly. We find
two qubit quantum pure states required for this purpose.
\end{abstract}



\section{Introduction}

Quantum game theory started from a seminal paper by Meyer \cite{Meyer}. Later
Eisert et. al. \cite{Eisert} studied the important bimatrix game of Prisoner's
Dilemma (PD) while focussing on the concept of Nash equilibrium (NE) from
noncooperative game theory. This concept attracted much attention in other
recent works in quantum game theory \cite{Eisert, Marinatto, Jianfeng,
Jianfeng1}. In fact Cournot (1838) \cite{cournot} anticipated Nash's
definition of equilibrium by over a century but only in the context of a
particular model of duopoly. In economics, an oligopoly is a form of market in
which a number $n$ of producers, say, $n\geq2$, and no others, provide the
market with a certain commodity. In the special case where $n=2$, it is called
a duopoly. Cournot's work \cite{cournot} is one of the classics of game theory
and also a cornerstone of the theory of industrial organization \cite{Tirole}.
In Cournot model of duopoly two-firms simultaneously put certain quantities of
a homogeneous product in the market. Cournot obtained an equilibrium value for
the quantities both firms will decide to put in the market. This equilibrium
value was based on a rule of behavior which says that if all the players
except one abide by it, the remaining player cannot do better than to abide by
it too. Nash gave a general concept of an equilibrium point in a
noncooperative game but existence of an equilibrium in duopoly game was known
much earlier. The ``Cournot equilibrium'' refers to NE in noncooperative form
of duopoly that Cournot considered.

In an interesting later development Stackelberg (1934) \cite{stackelberg,
gibbons} proposed a dynamic model of duopoly in which, contrary to Cournot's
assumption of simultaneous moves, a leader (or dominant) firm moves first and
a follower (or subordinate) firm moves second. A well known example is the
General Motors playing this leadership role in the early history of U.S.
automobile industry when more than one firms like Ford and Chrysler acted as
followers. In this sequential game a ``Stackelberg equilibrium'' is obtained
using the \textit{backwards-induction outcome} of the game. In fact
Stackelberg equilibrium refers to the sequential-move nature of the game and
this is a stronger solution concept than NE because sequential move games
sometimes have multiple NE , only one of which is associated with the
backwards-induction outcome of the game \cite{gibbons}.

The seminal paper by Meyer \cite{Meyer} has initiated the new field of quantum
games and has motivated many people to look at games from quantum
perspectives. Eisert et. al. \cite{Eisert} quantized the famous bimatrix game
of Prisoner's Dilemma (PD) and showed that dilemma disappears in quantum
world. They allowed players an access to a maximally entangled state that can
be generated from a system of two qubits. Each player then unitarily
manipulates a qubit in his possession. Players apply a unitary operator from a
particular subset of the general set of unitary operators that also forms a
group. Benjamin \cite{benjamin} later showed that if the players have access
to the set of general unitary operators then there is no NE in pure strategies
in the PD game. Benjamin implied that Eisert's set of unitary operators was
carefully chosen to generate a NE that has no classical counterpart.

Eisert et. al. \cite{Eisert} used an unentangling gate in their scheme to be
put before the quantum state is forwarded to measuring apparatus that
collapses the wave function and gives the payoffs. The unentangling gate in
Eisert's scheme ensured that classical game could be reproducible but it
motivated Marinatto \cite{L. Marinatto} to question the necessity of its
presence in the scheme to play a quantum game. Marinatto and Weber
\cite{Marinatto} came up with a solution where an initial quantum state, they
called it an ``initial strategy'', is made available to the players. This
initial strategy is then unitarily manipulated by the players in the
``tactics'' phase of the game that consisted of applying two unitary and
Hermitian operators (the identity and an inversion operator) with classical
probabilities on the initial quantum strategy. This inversion operator reverts
the quantum state just like the Pauli's spin flip operator does
\cite{benjamin1}. Marinatto and Weber \cite{Marinatto} showed that for an
initial strategy that is a maximally entangled state a unique NE for the game
of Battle of Sexes can be obtained. Later in an interesting comment Benjamin
\cite{benjamin1} considered the players' access to apply only two unitary and
Hermitian operators with classical probabilities on a quantum state a severe
restriction on all quantum mechanically possible manipulations. Marinatto and
Weber replied \cite{MarinattoRep} that the only restriction on a quantum form
of a game is that corresponding classical game must be reproducible as a
special case of the quantum form.

Agreeing with this reply we studied the concept of evolutionary stability in
asymmetric as well as symmetric quantum games \cite{iqbal, iqbal1}. Our prime
motivation was an important and interesting element in Marinatto and Weber's
scheme. It was the fact that a switch-over between a classical and a
non-classical form of a game could be achieved by having a control over the
parameters of the initial quantum state or initial strategy. In the scheme of
Eisert, Wilkens, and Lewenstein \cite{Eisert} such a switch-over takes place
when the players assign specific values to the parameters of the unitary
operators in their possession; while the initial two-qubit quantum state
always remains to be maximally entangled. Starting a quantum game from a
general pure state of two qubits that have interacted previously we were able
to show the possibility that an unstable symmetric NE of a bimatrix classical
game becomes stable in a quantum form of the same game \cite{iqbal2}. Our
assumption in this approach was that a replacement of the maximally entangled
two qubit quantum state, that Marinatto and Weber used to get a unique NE,
with a general two qubit pure quantum state also results in another form of
the same game. This assumption led us to find the conditions on the constants
of a general three qubit pure state that made it possible to counter against
coalition formation in a three player symmetric game \cite{iqbal3}. The person
responsible for preparing quantum states can thus make vanish the motivation
to make a coalition in the cooperative game.

Motivated by these recent developments in quantum games as well as the notion
of backwards-induction outcome of a dynamic game of complete information
\cite{gibbons} we present a quantum perspective on the interesting game of
Stackelberg duopoly. In present paper we start with the same assumption that a
game is decided only by players' unitary manipulations, payoff operators, and
the measuring apparatus deciding payoffs. When these are same a different
input quantum initial state gives only a different form of the same game. This
was our assumption when we studied evolutionary stability of a mixed NE in
Rock-Scissors-Paper game \cite{iqbal2}. Therefore all the games that can be
obtained by using a general two qubit pure state are only different forms of
the same game if the rest of the procedures in playing the quantum game are
same. For example in Marinatto and Weber's Battle of Sexes \cite{Marinatto}
the game remains Battle of Sexes for all two qubit pure quantum states.

With this assumption we start an analysis of Stackelberg duopoly by asking a
fundamental question: Is it possible to find a two qubit pure quantum state
that generates the classical Cournot equilibrium as a backwards-induction
outcome of the quantum form of Stackelberg duopoly? Why this question can be
of interest? For us it is interesting because in case the answer is yes then
the very important resource in quantum game theory i.e. entanglement can
potentially be a particularly useful element for `follower' in the
leader-follower model of Stackelberg duopoly \cite{gibbons}. This is because
in classical settings when static duopoly changes itself into a dynamic form
the follower becomes worse-off compared to leader who becomes better-off. We
find that under certain restrictions it is possible to find the needed two
qubit quantum states. Therefore a quantum form of a dynamic game of complete
information gives out an equilibrium that corresponds to classical static form
of the same game. In our analysis the equilibrium can be obtained for a
certain range of the constant of the duopoly game. This restriction appears
only due to an assumption that we introduce to simplify calculations. This
fact, however, does not rule out the possibility of getting a quantum form of
Stackelberg duopoly game with no such restriction.

\section{Backwards-induction outcome}

To make this paper self-contained we give an introduction of the
backwards-induction outcome of a sequential game. For this we find very useful
the ref. \cite{gibbons}. Consider a simple three step game

\begin{enumerate}
\item  Player 1 chooses an action $a_{1}$ from the feasible set $A_{1}$

\item  Player 2 observes $a_{1}$ and then chooses an action $a_{2}$ from the
feasible set $A_{2}$

\item  Payoffs are $u_{1}(a_{1},a_{2})$ and $u_{2}(a_{1},a_{2})$
\end{enumerate}

This game is an example of the dynamic games of complete and perfect
information. Key features of such games are

\begin{enumerate}
\item  the moves occur in sequence

\item  all previous moves are known before next move is chosen, and

\item  the players' payoffs are common knowledge. At the second stage of the
game when player $2$ gets the move he or she faces the following problem,
given the action $a_{1}$previously chosen by the player
\end{enumerate}%

\begin{equation}
\underset{a_{2}\in A_{2}}{Max}u_{2}(a_{1},a_{2}) \label{max2}%
\end{equation}
Assume that for each $a_{1}$ in $A_{1}$, player $2$'s optimization problem has
a unique solution $R_{2}(a_{1})$ also called the best response of player $2$.
Now player $1$ can also solve player $2$'s optimization problem by
anticipating player $2$'s response to each action $a_{1}$ that player $1$
might take. So that player $1$ faces the following problem%

\begin{equation}
\underset{a_{1}\in A_{1}}{Max}u_{1}(a_{1},R_{2}(a_{1})) \label{max1}%
\end{equation}
Suppose this optimization problem also has a unique solution for player $1$
and is denoted by $a_{1}^{\star}$. The solution $(a_{1}^{\star},R_{2}%
(a_{1}^{\star}))$ is the backwards-induction outcome of this game. In a simple
version of the Cournot's model two firms simultaneously decide the quantities
$q_{1}$ and $q_{2}$ respectively of a homogeneous product they want to put
into the market. Suppose $Q$ is the aggregate quantity i.e. $Q=q_{1}+q_{2}$
and $P(Q)=a-Q$ be the market-clearing price, the price at which all products
or services available in a market will find buyers. Assume the total cost to a
firm producing quantity $q_{i}$ is $cq_{i}$ i.e. there are no fixed costs and
the marginal cost is a constant $c$ with $c<a$. In a two-player game
theoretical model of this situation a firm's payoff or profit can be written
as \cite{gibbons}%

\begin{equation}
P_{i}(q_{i},q_{j})=q_{i}\left[  P(Q)-c\right]  =q_{i}\left[  a-c-(q_{i}%
+q_{j})\right]  =q_{i}\left[  k-(q_{i}+q_{j})\right]  \label{PayoffEq}%
\end{equation}
Solving for the NE easily gives the Cournot equilibrium%

\begin{equation}
q_{1}^{\star}=q_{2}^{\star}=\frac{k}{3} \label{Ceqbrm}%
\end{equation}
At this equilibrium the payoffs to both firms from eq. (\ref{PayoffEq}) are%

\begin{equation}
P_{1}(q_{1}^{\star},q_{2}^{\star})_{Cournot}=P_{2}(q_{1}^{\star},q_{2}^{\star
})_{Cournot}=\frac{k^{2}}{9} \label{CourPayoffs}%
\end{equation}
We now come to consider the classical form of duopoly game when it becomes
dynamic. The game becomes dynamic but the payoffs to players are given by the
same eq. (\ref{PayoffEq}) as for the case of the Cournot's game. We find
backwards-induction outcome in classical and a quantum form of Stackelberg's
duopoly. Taking advantage from bigger picture given to this dynamic game by
Hilbert space structure we then find two qubit pure quantum states that give
classical Cournot's equilibrium as the backwards-induction outcome of the
quantum game of Stackelberg's duopoly.

\section{Stackelberg duopoly}

\subsection{Classical form}

A leader (or dominant) firm moves first and a follower (or subordinate) firm
moves second in Stackelberg model of duopoly \cite{gibbons}. The sequence of
events is

\begin{enumerate}
\item  firm $A$ chooses a quantity $q_{1}\geq0$

\item  firm $B$ observes $q_{1}$ and then chooses a quantity $q_{2}\geq0$

\item  the payoffs to firms $A$ and $B$ are given by their respective profit
functions as
\end{enumerate}%

\begin{align}
P_{A}(q_{1},q_{2})  &  =q_{1}\left[  k-(q_{1}+q_{2})\right] \nonumber\\
P_{B}(q_{1},q_{2})  &  =q_{2}\left[  k-(q_{1}+q_{2})\right]  \label{ProfFunc}%
\end{align}
The backwards-induction outcome is found by first finding firm $B$'s reaction
to an arbitrary quantity by firm $A.$ Denoting this quantity as $R_{2}(q_{1})$
we find%

\begin{equation}
R_{2}(q_{1})=\underset{q_{2}\geq0}{Max}P_{B}(q_{1},q_{2})=\frac{k-q_{1}}{2}
\label{bestRes}%
\end{equation}
with $q_{1}<k$. Now the interesting aspect of this game is that firm $A$ can
solve the firm $B$'s problem as well. Therefore firm $A$ can anticipate that a
choice of the quantity $q_{1}$ will meet a reaction $R_{2}(q_{1})$. In the
first stage of the game firm $A$ can then compute a solution to his/her
optimization problem as%

\[
\underset{q_{1}\geq0}{Max}P_{A}\left[  q_{1},R_{2}(q_{1})\right]
=\underset{q_{1}\geq0}{Max}\frac{q_{1}(k-q_{1})}{2}%
\]
It gives%

\begin{equation}
q_{1}^{\star}=\frac{k}{2}\text{ \ \ and \ \ }R_{2}(q_{1}^{\star})=\frac{k}{4}
\label{StkEq}%
\end{equation}
It is the classical backwards-induction outcome of dynamic form of the duopoly
game. At this equilibrium payoffs or profits to players $A$ and $B$ are given
by eqs. (\ref{ProfFunc}) and (\ref{StkEq})%

\begin{equation}
P_{A}\left[  q_{1}^{\star},R_{2}(q_{1}^{\star})\right]  _{Stackelberg}%
=\frac{k^{2}}{8},\text{ \ \ \ \ }P_{B}\left[  q_{1}^{\star},R_{2}(q_{1}%
^{\star})\right]  _{Stackelberg}=\frac{k^{2}}{16} \label{StkPayffs}%
\end{equation}
From eq. (\ref{StkPayffs}) find the following ratio%

\begin{equation}
\frac{P_{A}\left[  q_{1}^{\star},R_{2}(q_{1}^{\star})\right]  _{Stackelberg}%
}{P_{B}\left[  q_{1}^{\star},R_{2}(q_{1}^{\star})\right]  _{Stackelberg}}=2
\label{ratio}%
\end{equation}
showing that with comparison to Cournot game in Stackelberg game the leader
firm becomes better-off and the follower firm becomes worse-off. This aspect
also hints an important difference between single and multi-person decision
problems. In single-person decision theory having more information can never
make the decision maker worse-off. In game theory, however, having more
information (or, more precisely, having it made public that one has more
information) can make a player worse-off \cite{gibbons}.

Now we look at backwards-induction outcome in a quantum perspective. Our
motivation is an interesting aspect that quantum form can bring into
backwards-induction outcome. It is the possibility that the `extra
information' that firm $B$ has does not make firm $B$ worse-off.

\subsection{Quantum form}

Stackelberg duopoly is a two player sequential game. Meyer \cite{Meyer}
considered quantum form of sequential game of PQ Penny Flip by unitary
operations on single qubit. Important difference between Meyer's PQ Penny Flip
and Stackelberg duopoly is that at the second stage player in PQ Penny Flip
doesn't know the previous move but in Stackelberg duopoly he knows that.

Marinatto and Weber \cite{Marinatto} used two qubits to play the
noncooperative game of Battle of Sexes. Players apply unitary operators $I$
and $C$ with classical probabilities on a two-qubit pure quantum state. $I$ is
identity and $C$ is inversion or Pauli's spin-flip operator. We prefer this
scheme to play the sequential game of Stackelberg duopoly for two reasons:

\begin{enumerate}
\item  Interesting feature in this scheme that classical game is reproducible
on making initial state unentangled.

\item  We assumed that other games obtained from every pure two qubit initial
state are only quantum forms of the classical game provided players' actions
and payoff generating measurement are exactly same \cite{iqbal2}. This
assumption reduces the problem of finding a quantum form of Stackelberg
duopoly with property that its equilibrium is same as in Cournot's duopoly to
the problem of finding conditions on parameters of two-qubit pure quantum
state. If the conditions are realistic then the corresponding quantum game
will give Cournot's equilibrium as a backwards-induction outcome.
\end{enumerate}%

\begin{center}
\includegraphics[
height=3.2335in,
width=4.2255in
]%
{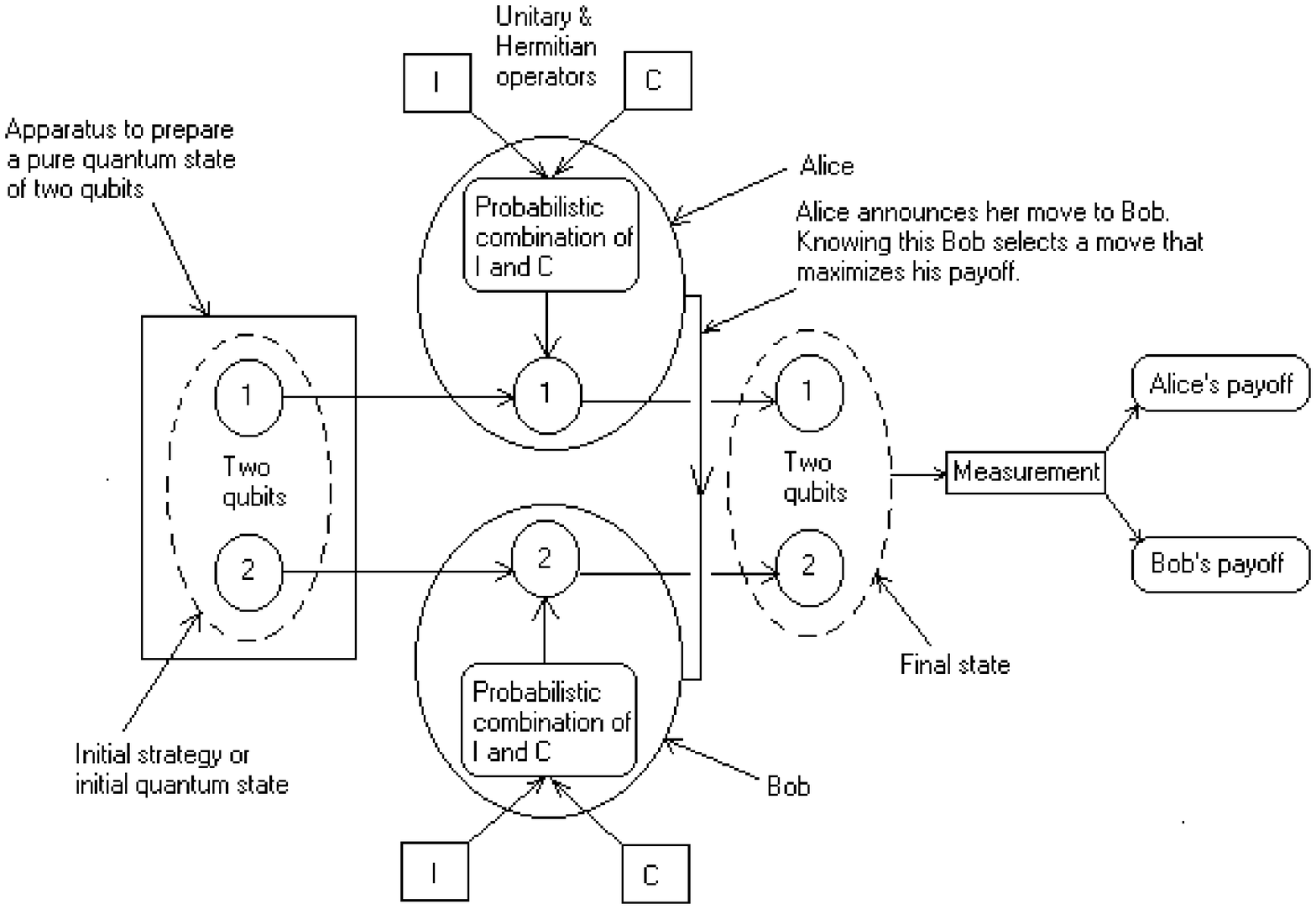}%
\\
Fig. 1. Playing a quantum form of Stackelberg duopoly.
\label{Fig. 1}%
\end{center}

Stackelberg duopoly is a dynamic game of complete information. Its quantum
form in Marinatto and Weber's scheme \cite{Marinatto} starts by preparing a
pure two-qubit initial quantum state. Both these qubits are then forwarded to
two players in the game called Alice and Bob. Marinatto and Weber expanded on
the earlier scheme proposed by Eisert, Wilkens, and Lewenstein \cite{Eisert}
using a two-qubit system in a maximally entangled quantum state. The
fundamental idea remains the same i.e. to exploit entanglement to get new
results for a simultaneous-move game. Suppose first move is to be played by
Alice and she plays her strategy by applying two operators in her possession
on her qubit with classical probabilities. She also announces her move
immediately so that Bob knows Alice's strategy before playing his move. Bob
plays his strategy on his qubit and both Alice and Bob forward their qubits to
a setup where measurement can be done to decide payoffs to both.

The concept of information about the previous moves is crucial for the
understanding of the present paper. A comparison between the sequential game
of Stackelberg duopoly with the simultaneous moves of the Battle of Sexes
makes clear the different information structure between the two games. For
example in the Battle of Sexes, when played sequentially, Alice does not
announce her first move to Bob before he makes his move. In this way the game
becomes sequential but the information structure of the game is still the same
as in its static form. Consequently, the sequential Battle of Sexes in the
above form has the same NE as in its static form. An unobserved-action form of
a game has the same NE as its simultaneous-move form. This observation led us
to play a quantum form of Stackelberg duopoly while keeping intact the
original structure of a scheme designed for simultaneous moves. A
consideration of playing a sequential game in a quantum way brings to mind the
Meyer's PQ Penny Flip \cite{Meyer} where only one qubit is used in the game.
Contrary to this, in present paper we use the two-qubit system of a
simultaneous moves game to play a sequential game.

One can ask what is the point of taking extra pains by using two qubits when a
quantum form of this sequential game can also be played by only one qubit in
similar way as Meyer's PQ Penny Flip. We prefer two qubits for a reason that
appeared to us quite important. \textit{In two-qubit case a comparison between
classical and a quantum form of the game translates itself into comparing two
games resulting from using unentangled and entangled initial quantum states}.
We do not rule out the possibility that a consideration of the dynamic game
using only single qubit gives equally or even more interesting results.

The quantum form of Marinatto and Weber's \cite{Marinatto} Battle of Sexes
game reduces to its classical form for the initial state $\left|  \psi
_{ini}\right\rangle =\left|  OO\right\rangle $ where $O$ represents the pure
classical strategy called `Opera'. In the same spirit we let classical payoffs
in Stackelberg duopoly given by eq. (\ref{ProfFunc}) reproduced when the
initial state $\left|  \psi_{ini}\right\rangle =\left|  11\right\rangle $ is
used to play the game. The state $\left|  11\right\rangle $ means that both
qubits are in lower state. We represent the upper state of a qubit by number
$2$. For $\left|  \psi_{ini}\right\rangle =\left|  11\right\rangle $ the
corresponding density matrix is%

\begin{equation}
\rho_{ini}=\left|  11\right\rangle \left\langle 11\right|  \label{IniDenMat}%
\end{equation}
When players apply the unitary operators $I$ and $C$ such that Alice and Bob
apply $I$ with probabilities $x$ and $y$ respectively the density matrix
(\ref{IniDenMat}) changes to \cite{Marinatto}%

\begin{align}
\rho_{fin}  &  =xyI_{A}\otimes I_{B}\rho_{ini}I_{A}^{\dagger}\otimes
I_{B}^{\dagger}+x(1-y)I_{A}\otimes C_{B}\rho_{ini}I_{A}^{\dagger}\otimes
C_{B}^{\dagger}+\nonumber\\
&  y(1-x)C_{A}\otimes I_{B}\rho_{ini}C_{A}^{\dagger}\otimes I_{B}^{\dagger
}+(1-x)(1-y)C_{A}\otimes C_{B}\rho_{ini}C_{A}^{\dagger}\otimes C_{B}^{\dagger
}\nonumber\\
&  \label{FinDenMat}%
\end{align}
where the inversion operator $C$ interchanges the vectors $\left|
1\right\rangle $ and $\left|  2\right\rangle $ i.e. $C\left|  1\right\rangle
=\left|  2\right\rangle ,$ $C\left|  2\right\rangle =\left|  1\right\rangle $
and $C^{\dagger}=C=C^{-1}$. We now assume that in Stackelberg duopoly also
players' moves are again given by probabilities lying in the range $\left[
0,1\right]  $. The moves by Alice and Bob in classical duopoly game are given
by quantities $q_{1}$ and $q_{2}$ where $q_{1},q_{2}\in\lbrack0,\infty)$. We
assume that Alice and Bob agree on a function that can uniquely define a real
positive number in the range $(0,1]$ for every quantity $q_{1},q_{2}$ in
$[0,\infty)$. A simple such function is $\frac{1}{1+q_{i}}$, so that Alice and
Bob find $x$ and $y$ respectively as%

\begin{equation}
x=\frac{1}{1+q_{1}},\text{ \ \ \ \ \ }y=\frac{1}{1+q_{2}} \label{DefFuns}%
\end{equation}
and use these real numbers as probabilities with which they apply the identity
operator $I$ on the quantum state at their disposal. With a substitution from
eqs. (\ref{IniDenMat},\ref{DefFuns}) the final density matrix (\ref{FinDenMat}%
) can be written as%

\begin{equation}
\rho_{fin}=\frac{1}{(1+q_{1})(1+q_{2})}\left[  \left|  11\right\rangle
\left\langle 11\right|  +q_{1}q_{2}\left|  22\right\rangle \left\langle
22\right|  +q_{1}\left|  21\right\rangle \left\langle 21\right|  +q_{2}\left|
12\right\rangle \left\langle 12\right|  \right]
\end{equation}
We now also suppose that in the measurement and payoffs finding phase the
quantities $q_{1}$ and $q_{2}$ are known to the `agent' doing this action. The
agent applies the payoff operators $(P_{A})_{oper}$, $(P_{B})_{oper}$ given as follows%

\begin{align}
(P_{A})_{oper}  &  =(1+q_{1})(1+q_{2})q_{1}\left[  k\left|  11\right\rangle
\left\langle 11\right|  -\left|  21\right\rangle \left\langle 21\right|
-\left|  12\right\rangle \left\langle 12\right|  \right] \nonumber\\
(P_{B})_{oper}  &  =(1+q_{1})(1+q_{2})q_{2}\left[  k\left|  11\right\rangle
\left\langle 11\right|  -\left|  21\right\rangle \left\langle 21\right|
-\left|  12\right\rangle \left\langle 12\right|  \right]  \label{PayOpers}%
\end{align}
Note that the classical payoffs of eq. (\ref{ProfFunc}) are reproduced with
the initial state $\left|  \psi_{ini}\right\rangle =\left|  11\right\rangle $
with the following trace operations%

\begin{align}
P_{A}(q_{1},q_{2})  &  =Trace\left[  (P_{A})_{oper}\rho_{fin}\right]
\nonumber\\
P_{B}(q_{1},q_{2})  &  =Trace\left[  (P_{B})_{oper}\rho_{fin}\right]
\label{PlayersPayoffs}%
\end{align}
A more general form of quantum duopoly can now be played by keeping the payoff
operators of eq. (\ref{PayOpers}) in the agent's possession and preparing a
general initial pure state of the following form%

\begin{gather}
\left|  \psi_{ini}\right\rangle =c_{11}\left|  11\right\rangle +c_{12}\left|
12\right\rangle +c_{21}\left|  21\right\rangle +c_{22}\left|  22\right\rangle
\label{iniStat}\\
\text{where \ \ }\left|  c_{11}\right|  ^{2}+\left|  c_{12}\right|
^{2}+\left|  c_{21}\right|  ^{2}+\left|  c_{22}\right|  ^{2}=1 \label{Normn}%
\end{gather}
where $c_{ij}$,for $i,j=1$ or $2,$ are complex numbers and $\left|
ij\right\rangle $ are orthonormal basis vectors in $2\otimes2$ dimensional
Hilbert space. The payoffs to Alice and Bob can now be obtained in this more
general quantum game from eqs. (\ref{PlayersPayoffs}) that use the same payoff
operators of eqs. (\ref{PayOpers}). The payoffs to Alice and Bob can now be
written as \cite{iqbal4}%

\begin{align}
\left[  P_{A}(q_{1},q_{2})\right]  _{qtm}  &  =\frac{(\omega_{11}+\omega
_{12}q_{2})+q_{1}(\omega_{21}+\omega_{22}q_{2})}{(1+q_{1})(1+q_{2}%
)}\nonumber\\
\left[  P_{B}(q_{1},q_{2})\right]  _{qtm}  &  =\frac{(\chi_{11}+\chi_{12}%
q_{2})+q_{1}(\chi_{21}+\chi_{22}q_{2})}{(1+q_{1})(1+q_{2})} \label{QPayoffs}%
\end{align}
where the subscript $qtm$ is for `quantum' and%

\begin{align}
\left[
\begin{array}
[c]{c}%
\omega_{11}\\
\omega_{12}\\
\omega_{21}\\
\omega_{22}%
\end{array}
\right]   &  =\left[
\begin{array}
[c]{cccc}%
\left|  c_{11}\right|  ^{2} & \left|  c_{12}\right|  ^{2} & \left|
c_{21}\right|  ^{2} & \left|  c_{22}\right|  ^{2}\\
\left|  c_{12}\right|  ^{2} & \left|  c_{11}\right|  ^{2} & \left|
c_{22}\right|  ^{2} & \left|  c_{21}\right|  ^{2}\\
\left|  c_{21}\right|  ^{2} & \left|  c_{22}\right|  ^{2} & \left|
c_{11}\right|  ^{2} & \left|  c_{12}\right|  ^{2}\\
\left|  c_{22}\right|  ^{2} & \left|  c_{21}\right|  ^{2} & \left|
c_{12}\right|  ^{2} & \left|  c_{11}\right|  ^{2}%
\end{array}
\right]  \left[
\begin{array}
[c]{c}%
kq_{1}(1+q_{1})(1+q_{2})\\
-q_{1}(1+q_{1})(1+q_{2})\\
-q_{1}(1+q_{1})(1+q_{2})\\
0
\end{array}
\right] \nonumber\\
\left[
\begin{array}
[c]{c}%
\chi_{11}\\
\chi_{12}\\
\chi_{21}\\
\chi_{22}%
\end{array}
\right]   &  =\left[
\begin{array}
[c]{cccc}%
\left|  c_{11}\right|  ^{2} & \left|  c_{12}\right|  ^{2} & \left|
c_{21}\right|  ^{2} & \left|  c_{22}\right|  ^{2}\\
\left|  c_{12}\right|  ^{2} & \left|  c_{11}\right|  ^{2} & \left|
c_{22}\right|  ^{2} & \left|  c_{21}\right|  ^{2}\\
\left|  c_{21}\right|  ^{2} & \left|  c_{22}\right|  ^{2} & \left|
c_{11}\right|  ^{2} & \left|  c_{12}\right|  ^{2}\\
\left|  c_{22}\right|  ^{2} & \left|  c_{21}\right|  ^{2} & \left|
c_{12}\right|  ^{2} & \left|  c_{11}\right|  ^{2}%
\end{array}
\right]  \left[
\begin{array}
[c]{c}%
kq_{2}(1+q_{1})(1+q_{2})\\
-q_{2}(1+q_{1})(1+q_{2})\\
-q_{2}(1+q_{1})(1+q_{2})\\
0
\end{array}
\right] \nonumber\\
&  \label{consts}%
\end{align}
The classical payoffs of duopoly game given in eqs. (\ref{ProfFunc}) are
recovered from the eqs. (\ref{QPayoffs}) when the initial quantum state
becomes unentangled and given by $\left|  \psi_{ini}\right\rangle =\left|
11\right\rangle $. Classical duopoly is, therefore, a subset of its quantum version.

We now find the backwards-induction outcome in this quantum form of
Stackelberg duopoly. We proceed in exactly the same way as it is done in the
classical game except that players' payoffs are now given by eqs.
(\ref{QPayoffs}) and not by eqs. (\ref{ProfFunc}). The first step in
backwards-induction in quantum game is to find Bob's reaction to an arbitrary
quantity $q_{1}$ chosen by Alice. Denoting this quantity as $\left[
R_{2}(q_{1})\right]  _{qtm}$ we find%

\begin{gather}
\left[  R_{2}(q_{1})\right]  _{qtm}=\underset{q_{2}\geq0}{Max}\left[
P_{B}(q_{1},q_{2})\right]  _{qtm}=\frac{q_{1}\triangle_{1}+\triangle_{2}%
}{-2\left\{  q_{1}\triangle_{3}+\triangle_{4}\right\}  }\text{ \ \ \ \ where}%
\nonumber\\
\left|  c_{11}\right|  ^{2}+\left|  c_{22}\right|  ^{2}-k\left|
c_{21}\right|  ^{2}=\triangle_{1}\text{, \ \ \ \ \ \ \ \ }\left|
c_{12}\right|  ^{2}+\left|  c_{21}\right|  ^{2}-k\left|  c_{11}\right|
^{2}=\triangle_{2}\nonumber\\
\left|  c_{12}\right|  ^{2}+\left|  c_{21}\right|  ^{2}-k\left|
c_{22}\right|  ^{2}=\triangle_{3}\text{, \ \ \ \ \ \ \ \ }\left|
c_{11}\right|  ^{2}+\left|  c_{22}\right|  ^{2}-k\left|  c_{12}\right|
^{2}=\triangle_{4} \label{QbestRes}%
\end{gather}
This reaction reduces to its classical value of eq. (\ref{bestRes}) when
$\left|  c_{11}\right|  ^{2}=1$. Similar to classical game Alice can now solve
Bob's problem as well. Therefore Alice can anticipate that a choice of the
quantity $q_{1}$ will meet a reaction $\left[  R_{2}(q_{1})\right]  _{qtm}$.
In the first stage of the game like its classical version Alice can compute a
solution to her optimization problem as%

\begin{equation}
\underset{q_{1}\geq0}{Max}\left[  P_{A}\left\{  q_{1},\left\{  R_{2}%
(q_{1})\right\}  _{qtm}\right\}  \right]  _{qtm}%
\end{equation}
To find it Alice calculates the following quantity%

\begin{align}
\frac{d\left[  P_{A}(q_{1},q_{2})\right]  _{qtm}}{dq_{1}}  &  =\frac{(\left|
c_{11}\right|  ^{2}+\left|  c_{22}\right|  ^{2}-\left|  c_{12}\right|
^{2}-\left|  c_{21}\right|  ^{2})}{(1+q_{1})}\left\{  -2q_{1}^{2}%
+q_{1}(k-2)+k\right\} \nonumber\\
&  +(1+2q_{1})\left\{  (k-1)\left|  c_{21}\right|  ^{2}-\left|  c_{12}\right|
^{2}\right\}  +k(\left|  c_{12}\right|  ^{2}-\left|  c_{22}\right|
^{2})\nonumber\\
&  -q_{1}\frac{dq_{2}}{dq_{1}}\left\{  \triangle_{4}+q_{1}\triangle
_{3}\right\}  -q_{2}\left\{  2q_{1}\triangle_{3}+\triangle_{4}\right\}
\nonumber\\
&  \label{derivative}%
\end{align}
and replaces $q_{2}$ in eq. (\ref{derivative}) with $\left[  R_{2}%
(q_{1})\right]  _{qtm}$ given by eq. (\ref{QbestRes}) and then equates eq.
(\ref{derivative}) to zero to find a $q_{1}^{\star}$ that maximizes her payoff
$\left[  P_{A}(q_{1},q_{2})\right]  _{qtm}$. For a maxima she would ensure
that the second derivative of $P_{A}\left\{  q_{1},\left\{  R_{2}%
(q_{1})\right\}  _{qtm}\right\}  $ with respect to $q_{1}$ at $q_{1}%
=q_{1}^{\star}$ is a negative quantity. The $q_{1}^{\star}$ together with
$\left[  R_{2}(q_{1}^{\star})\right]  _{qtm}$ will form the
backwards-induction outcome of the quantum game.

An interesting situation is when the backwards-induction outcome in quantum
version of Stackelberg duopoly becomes same as the classical Cournot
equilibrium of duopoly. The classical situation of leader becoming better-off
while the follower becoming worse-off is then avoided in the quantum form of
Stackelberg duopoly. To look for this possibility we need such an initial
state $\left|  \psi_{ini}\right\rangle =c_{11}\left|  11\right\rangle
+c_{12}\left|  12\right\rangle +c_{21}\left|  21\right\rangle +c_{22}\left|
22\right\rangle $ that at $q_{1}^{\star}=q_{2}^{\star}=\frac{k}{3}$ we should
have following relations holding true also with the normalization condition
given in eq. (\ref{Normn})%

\begin{gather}
\frac{d\left[  P_{A}\left\{  q_{1},\left\{  R_{2}(q_{1})\right\}
_{qtm}\right\}  \right]  _{qtm}}{dq_{1}}\mid_{q_{1}=q_{1}^{\star}%
}=0\label{Condn1}\\
\left[  \frac{d^{2}\left[  P_{A}\left\{  q_{1},\left\{  R_{2}(q_{1})\right\}
_{qtm}\right\}  \right]  _{qtm}}{dq_{1}^{2}}\mid_{q_{1}=q_{1}^{\star}}\right]
<0\label{Condn2}\\
q_{2}^{\star}=\left\{  R_{2}(q_{1}^{\star})\right\}  _{qtm} \label{Condn3}%
\end{gather}
The conditions (\ref{Condn1},\ref{Condn2}) simply say that the
backwards-induction outcome of the quantum game is the same as Cournot
equilibrium in classical game. The condition (\ref{Condn3}) says that Bob's
reaction to Alice's choice of $q_{1}^{\star}=\frac{k}{3}$ is $q_{2}^{\star
}=\frac{k}{3}$. To show that such quantum states can exist for which the
conditions (\ref{Condn1},\ref{Condn2},\ref{Condn3}) with the normalization
(\ref{Normn}) hold true we give an example where $\left|  c_{11}\right|
^{2},\left|  c_{12}\right|  ^{2}$ and $\left|  c_{21}\right|  ^{2}$ are
written as functions of $k$ with our assumption that $\left|  c_{22}\right|
^{2}=0$. Though this assumption puts its own restriction on the possible range
of $k$ for which the above conditions hold for these functions but still it
shows clearly the possibility of finding the required initial quantum states.
The functions are found as%

\begin{align*}
\left|  c_{12}(k)\right|  ^{2}  &  =\frac{-f(k)+\sqrt{f(k)^{2}-4g(k)h(k)}%
}{2g(k)}\text{ \ \ where}\\
f(k)  &  =j(k)\left\{  \frac{-7}{18}k^{2}+\frac{1}{3}k+\frac{1}{2}\right\}
+\left\{  \frac{k^{2}}{9}+\frac{k}{3}+\frac{1}{2}\right\} \\
g(k)  &  =j(k)^{2}\left\{  \frac{-1}{9}k^{3}+\frac{7}{18}k^{2}-\frac{1}%
{2}\right\}  +j(k)\left\{  \frac{2}{9}k^{3}+\frac{5}{18}k^{2}-\frac{1}%
{2}k-1\right\}  +\\
&  \left\{  \frac{-1}{9}k^{2}-\frac{1}{2}k-\frac{1}{2}\right\} \\
h(k)  &  =\frac{-1}{6}k\text{, \ \ \ \ \ \ }j(k)=\frac{9-4k^{2}}{k^{2}-9}%
\end{align*}
also%

\begin{align}
\left|  c_{21}(k)\right|  ^{2}  &  =j(k)\left|  c_{12}(k)\right|  ^{2}\\
\left|  c_{11}(k)\right|  ^{2}  &  =1-\left|  c_{12}(k)\right|  ^{2}-\left|
c_{21}(k)\right|  ^{2}%
\end{align}
Now, interestingly, given that allowed range of $k$ is $1.5\leq k\leq1.73205
$, all of the conditions (\ref{Normn},\ref{Condn1},\ref{Condn2},\ref{Condn3})
hold at $q_{1}^{\star}=q_{2}^{\star}=\frac{k}{3}$. So that in this range of
$k$ a quantum form of Stackelberg duopoly exists that gives the classical
Cournot equilibrium as backwards-induction outcome. The restriction on allowed
range of $k$ is result of our assumption that $\left|  c_{22}(k)\right|
^{2}=0$. In fig.$2$ below $\left|  c_{11}(k)\right|  ^{2},\left|
c_{12}(k)\right|  ^{2}$and $\left|  c_{21}(k)\right|  ^{2}$ are plotted
against $k$ in the above range.%

\begin{center}
\includegraphics[
height=2.8997in,
width=4.4304in
]%
{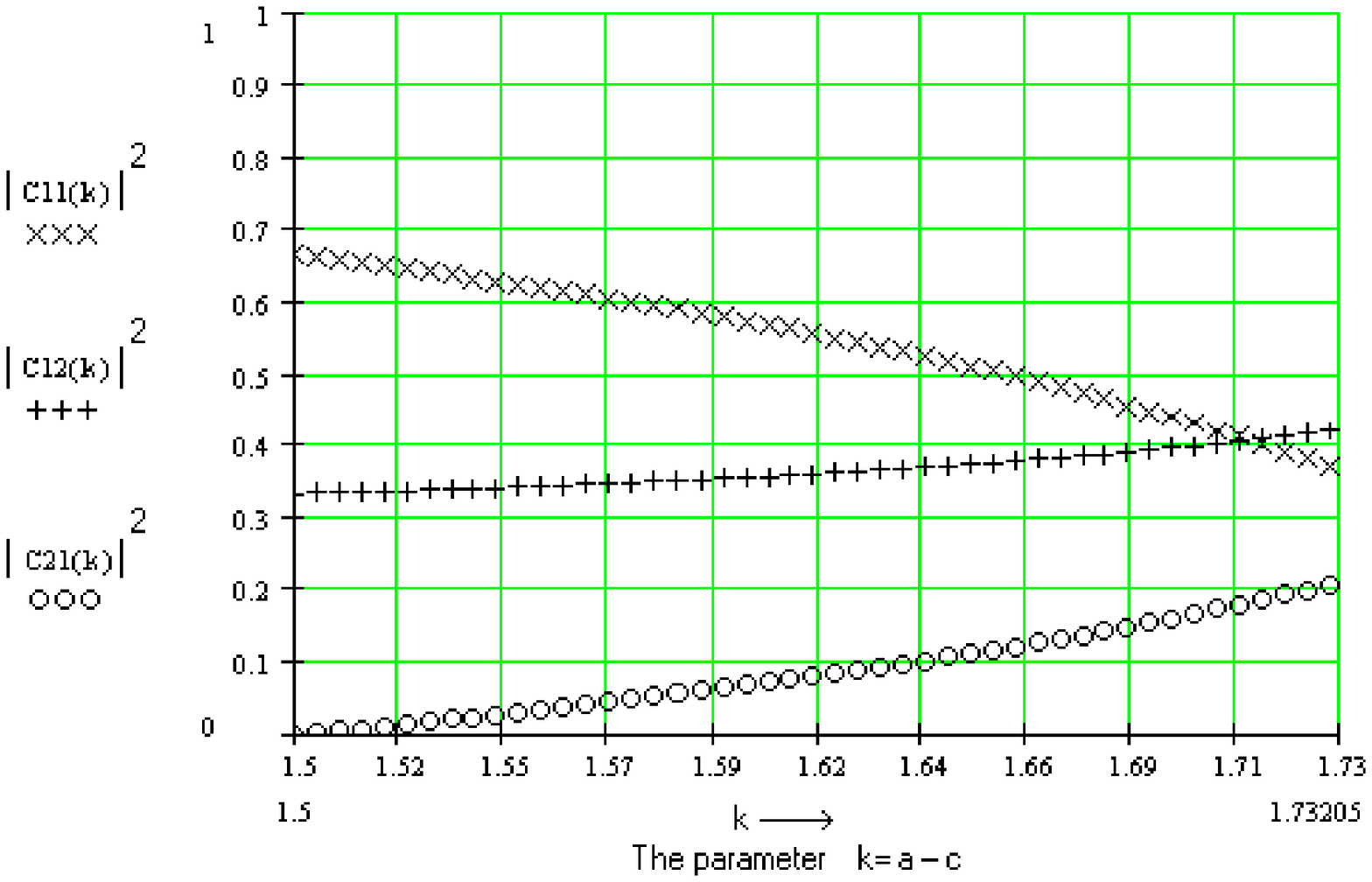}%
\\
Fig. 2. $\left|  c_{11}(k)\right|  ^{2},\left|  c_{12}(k)\right|  ^{2}\&$
$\left|  c_{21}(k)\right|  ^{2}$against $k$ when $\left|  c_{22}\right|
^{2}=0$
\label{Fig. 2}%
\end{center}

\section{Discussion and Conclusion}

What can possibly be a relevance of considering a quantum form of a game that
models a competition between two firms in macroscopic world of economics?
Quantum mechanics was developed to understand phenomena in the regime of
atomic and subatomic interactions and is still mostly used in that domain.
What is of interest in extending a game theoretical model of interaction
between firms towards quantum domain? These questions naturally arise not only
with reference to Stackelberg duopoly considered in this paper but also other
related works in quantum games. Apart from exciting new directions that
quantum mechanics brings to game theory there is also a fundamental interest
in quantum games from the view of quantum information theory. The fact that
quantum algorithms may be thought of as games between classical and quantum
agents was pointed out by Meyer \cite{Meyer}. Meyer indicated a strong
motivation for the study of quantum games by considering zero-sum quantum
games in order to have a new starting point at hand to find quantum algorithms
that outperform its classical analogue. Many quantum information exchange
protocols have been modelled like games. Eavesdropping and optimal cloning are
two such often cited examples where objective before a player is to gain as
much information as possible. We believe that like other notions of game
theory finding some relevance in quantum information a consideration of
backwards-induction can be of interest for exactly the same reasons. It does
not seem hard to imagine situations in quantum information where moves occur
in sequence, all previous moves are observed before the next move is chosen,
players' payoffs from each feasible combination of moves are common knowledge.
Interesting questions then arise about how a quantum version of dynamic game
of complete information can influence the outcome. Our primary motivation,
however, to study backwards-induction in quantum games is from the view of
dynamic stability, especially of symmetric NE, that has important relevance in
evolutionary games \cite{cressman} that we found interesting in quantum
settings \cite{iqbal, iqbal1, iqbal2, iqbal4}.

The duopoly game models economic competition between firms and applied
economics is the area where it is studied in detail. We considered this game
in a scheme that tells how to play a quantum game and gives a Hilbert
structure to the strategy space to which players have access to. The fact that
quantum game theory can give entirely new views on games important in
economics is apparent in recent interesting papers by Piotrowski and
Sladkowski \cite{piotrowski, piotrowski1} proposing a quantum-like description
of markets and economies where players' strategies belong to Hilbert space. It
shows that quantum games certainly have features of interest to applied
economists. Reciprocating with it we showed that games played by firms in
economic competition can give counter-intuitive solutions when played in a
quantum world.

We conclude our results as follows. A comparison between the NE in Cournot
game with backwards-induction outcome in classical Stackelberg duopoly shows
that having Alice (or firm $A$ who acts first) know that Bob (or firm $B$ who
acts second) knows $q_{1}$ (Alice's move) hurts Bob. In fact in classical
Stackelberg game Bob should not believe that Alice has chosen its Stackelberg
quantity i.e. $q_{1}^{\star}=\frac{k}{2}$. We have shown that there can be a
quantum version of Stackelberg duopoly where Bob is not hurt even if he knows
the quantity $q_{1}$ chosen by Alice. The backwards-induction outcome of this
quantum game is the same as the NE in classical Cournot game where decisions
are made simultaneously and there is no such information that hurts a player.
Though this outcome in quantum game is obtained for a restricted range of the
variable $k$ but it is only because of a simplification in calculations.

\section{Acknowledgment}

This work is supported by Pakistan Institute of Lasers and Optics, Islamabad.


\begin{thebibliography}{99}
\bibitem{Meyer}D. A. Meyer. Quantum Strategies. Phy. Rev. Lett. 82, 1052-1055
(1999). quant-ph/9804010. Also D. A. Meyer. Quantum games and quantum
algorithms. quant-ph/0004092

\bibitem {Eisert}J. Eisert, M. Wilkens, M. Lewenstein. Quantum Games and
Quantum Strategies. Phys. Rev. Lett. 83, 3077 (1999). quant-ph/9806088. Also
J. Eisert, M. Wilkens. Quantum Games. J. Mod. Opt. 47 (2000) 2543. quant-ph/0004076

\bibitem {Marinatto}L. Marinatto, T. Weber. A quantum approach to static games
of complete information. Phy. Lett. A 272, 291 (2000). quant-ph/0004081

\bibitem {Jianfeng}Jiangfeng Du, Xiaodong Xu, Hui Li, Xianyi Zhou, Rongdian
Han. Nash Equilibrium in the Quantum Battle of Sexes Game. quant-ph/0010050

\bibitem {Jianfeng1}Jiangfeng Du, Hui Li, Xiaodong Xu, Mingjun Shi, Xianyi
Zhou, Rongdian Han. Remark On Quantum Battle of The Sexes Game. quant-ph/0103004

\bibitem {cournot}A. Cournot. \textit{Researches into the Mathematical
Principles of the Theory of Wealth}. Edited by N. Bacon. New York: Macmillan, 1897

\bibitem {gibbons}R. Gibbons. \textit{Game Theory for Applied Economists}.
Princeton University Press. 1992

\bibitem {Tirole}J. Tirole. \textit{The theory of industrial organization}.
Cambridge: MIT Press. 1988

\bibitem {stackelberg}H. von Stackelberg. \textit{Marktform und
Gleichgewicht.} Vienna: Julius Springer. 1934

\bibitem {benjamin}S. C. Benjamin, P. M. Hayden. Comment on `Quantum Games and
Quantum Strategies' quant-ph/0003036

\bibitem {benjamin1}S. C. Benjamin. Comment on: ''A quantum approach to static
games of complete information''. quant-ph/0008127

\bibitem {benjamin2}S. C. Benjamin, P. M. Hayden. Multi-Player Quantum Games. quant-ph/0007038

\bibitem {L. Marinatto}L. Marinatto. Private communication.

\bibitem {MarinattoRep}L. Marinatto, T. Weber. Reply to ''Comment on: A
Quantum Approach to Static Games of Complete Information''. Physics Letters A
277, 183-184 (2000). quant-ph/0009103

\bibitem {cressman}R. Cressman. The Dynamic (In)Stability of Backwards
Induction. Journal of Economic Theory. 83, 260-285 (1998)

\bibitem {iqbal}A. Iqbal and A. H. Toor. Evolutionarily stable strategies in
quantum games. Phys. Lett. A 280 (2001) 249-256. quant-ph/0007100

\bibitem {iqbal1}A. Iqbal and A. H. Toor. Entanglement and Dynamic Stability
of Nash Equilibria in a Symmetric Quantum Game. Phys. Lett. A 286 (2001)
245-250. quant-ph/0101106

\bibitem {iqbal2}A. Iqbal and A. H. Toor. Quantum Mechanics gives Stability to
a Nash Equilibrium. Phys. Rev. A 65, 022306 (2002). quant-ph/0104091

\bibitem {iqbal3}A. Iqbal and A. H. Toor. Quantum Cooperative Games. Phys.
Lett. A 293/3-4 (2002) 103-108. quant-ph/0108091

\bibitem {iqbal4}A. Iqbal and A. H. Toor. Evolutionary stability of mixed Nash
equilibrium in quantized symmetric bimatrix games. quant-ph/0106056

\bibitem {piotrowski}E. W. Piotrowski, J. Sladkowski. Quantum Market Games. quant-ph/0104006

\bibitem {piotrowski1}E. W. Piotrowski, J. Sladkowski. Quantum Bargaining
Games. quant-ph/0106140
\end{thebibliography}
\end{document}